\documentclass[showpacs,twocolumn,groupedaddress,prl,aps,%
  longbibliography,superscriptaddress]{revtex4-1}

\usepackage{amsmath}
\usepackage{amssymb}
\usepackage{maybemath}
\usepackage{xcolor}
\usepackage{graphicx}
\usepackage{bm}
\usepackage{comment}
\usepackage{braket}
\usepackage{mathrsfs}
\usepackage{tikz}

\usetikzlibrary{decorations.pathmorphing}
\usetikzlibrary{shapes}
\usetikzlibrary{shapes.geometric}

\tikzset{
    magnetic/.style={
        fill,
        shape border rotate=-90,
        isosceles triangle,
        isosceles triangle apex angle=60,
        node distance=1,
        minimum height=.1
    }
}

\tikzset{
    othermagnetic/.style={
        fill,
        shape border rotate=90,
        isosceles triangle,
        isosceles triangle apex angle=60,
        node distance=1,
        minimum height=.1
    }
}

\tikzset{
    leftmagnetic/.style={
        fill,
        shape border rotate=0,
        isosceles triangle,
        isosceles triangle apex angle=60,
        node distance=1,
        minimum height=.1
    }
}

\tikzset{
    rightmagnetic/.style={
        fill,
        shape border rotate=180,
        isosceles triangle,
        isosceles triangle apex angle=60,
        node distance=1,
        minimum height=.1
    }
}

\pgfdeclaredecoration{complete sines}{initial}
{
    \state{initial}[
        width=+0pt,
        next state=upsine,
        persistent precomputation={\pgfmathsetmacro\matchinglength{
            \pgfdecoratedinputsegmentlength / int(\pgfdecoratedinputsegmentlength/\pgfdecorationsegmentlength)}
            \setlength{\pgfdecorationsegmentlength}{\matchinglength pt}
        }] {}
    \state{upsine}[width=\pgfdecorationsegmentlength,next state=downsine]{
        \pgfpathsine{\pgfpoint{0.25\pgfdecorationsegmentlength}{0.5\pgfdecorationsegmentamplitude}}
        \pgfpathcosine{\pgfpoint{0.25\pgfdecorationsegmentlength}{-0.5\pgfdecorationsegmentamplitude}}
    }
    \state{downsine}[width=\pgfdecorationsegmentlength,next state=upsine]{
        \pgfpathsine{\pgfpoint{0.25\pgfdecorationsegmentlength}{-0.5\pgfdecorationsegmentamplitude}}
        \pgfpathcosine{\pgfpoint{0.25\pgfdecorationsegmentlength}{0.5\pgfdecorationsegmentamplitude}}
}
    \state{final}{}
}

\usepackage{dcolumn}
\newcolumntype{.}{D{.}{.}{-1}}

\usepackage[%
  colorlinks=true,
  urlcolor=blue,
  linkcolor=blue,
  citecolor=blue
]{hyperref}

\makeatletter

\newcommand{\Rmnum}[1]{\expandafter\@slowromancap\romannumeral #1@}
\makeatother

\begin{document}

\newcommand{\addrMPIK}{Max Planck Institute for Nuclear Physics, Saupfercheckweg 1, 69117 Heidelberg, Germany}

\title{Two-loop virtual light-by-light scattering corrections to the bound-electron \boldmath{$g$} factor}

\author{V. Debierre}
\email[]{vincent.debierre@mpi-hd.mpg.de}
\author{B. Sikora}
\author{H. Cakir}
\author{N.~S. Oreshkina}
\affiliation{\addrMPIK}
\author{V.~A. Yerokhin}
\affiliation{\addrMPIK}
\affiliation{Center for Advanced Studies, Peter the Great St. Petersburg Polytechnic University, 195251 St. Petersburg, Russia}
\author{C.~H. Keitel}
\author{Z. Harman}
\affiliation{\addrMPIK}

\begin{abstract}
A critical set of two-loop QED corrections to the $g$ factor of hydrogenlike ions is calculated without expansion in the nuclear binding field. These corrections are due to the polarization of the external magnetic field by the quantum vacuum, which is dressed by the binding field. The result obtained for the self-energy--magnetic-loop diagrams is compared with the current state-of-the-art result, derived through a perturbative expansion in the binding strength parameter $Z\alpha$, with $Z$ the atomic number and $\alpha$ the fine-structure constant. Agreement is found in the $Z\rightarrow0$ limit. However, even for very light ions, the perturbative result fails to approximate the magnitude of the corresponding correction to the $g$ factor. The total correction to the $g$ factor coming from all diagrams considered in this work is found to be highly relevant for upcoming experimental tests of fundamental physics with highly charged ions.
\end{abstract}


\maketitle

\textit{Introduction.---} Measurements of the $g$ factor of heavy hydrogenlike ions are projected at different facilities, such as the ALPHATRAP Penning trap~\cite{Sturm17,Sturm19,ArapoglouAcc} and the HITRAP facility~\cite{Qui01,Herfurth_2015,Vogel2019}. These measurements are forecast to match the most precise measurements of the $g$ factor so far~\cite{Sturm13,Sturm14}, which have an uncertainty of the order of $10^{-11}$. High-precision calculations and measurements of the bound-electron $g$ factor can be combined to perform state-of-the-art determinations of fundamental constants such as the electron mass $m_e$~\cite{Sturm14} and the fine-structure constant $\alpha$~\cite{WDiffOld,GFactorAlpha,HalilReduced}. Furthermore, heavy ions are an ideal testing ground for quantum electrodynamics (QED) calculations in the presence of strong fields~\cite{Sturm11,Sturm13,ShabaevReview,ConfReview,IndelicatoReview}, and measurements of their $g$ factor were recently shown to be a promising avenue in the search for physics beyond the Standard Model~\cite{FifthForceG}.

The interpretation of upcoming experiments on heavy hydrogenlike ions demands improvements in the theory, especially concerning the calculation of radiative corrections to the $g$ factor. The one-loop radiative corrections have been calculated nonperturbatively~\cite{BeierOneLoop,OneLoopGUehling,YeroGFactor,DelbrueckGArticle,LeeDelbrueck} in the electromagnetic binding parameter $Z\alpha$, but the calculation of the two-loop corrections has only been completed through orders $\left(Z\alpha\right)^4$~\cite{EveryoneTwoLoop,CzarneckiSzafron} and $\left(Z\alpha\right)^5$~\cite{CzarneckiSzafron,CzarneckiLetter}. Further progress of the two-loop calculation should be sought in the non-perturbative approach, especially for application to heavy ions. The non-perturbative evaluation of all twenty-nine non-equivalent two-loop diagrams contributing to the $g$ factor of a bound electron is one of the great challenges of present-day atomic QED theory. The results presented in this work constitute an important step towards the completion of this project.

A few years ago, the two-loop diagrams featuring two vacuum polarization (VP) loops, as well as those featuring one VP loop and one self-energy (SE) loop, were calculated~\cite{VladZVP} in the free-fermion loop-approximation. The challenging SE-SE diagrams are currently being computed~\cite{BastianPhD,SikoraTwoLoop}. Several VP-VP and VP-SE diagrams were not calculated in Ref.~\cite{VladZVP}, because they vanish in the free VP loop approximation. In this work, we go beyond this approximation and calculate these diagrams, and show that they must be taken into account for heavy ions at the current level of experimental accuracy. We also show that a perturbative calculation of the diagrams examined here is insufficient even for very light ions.

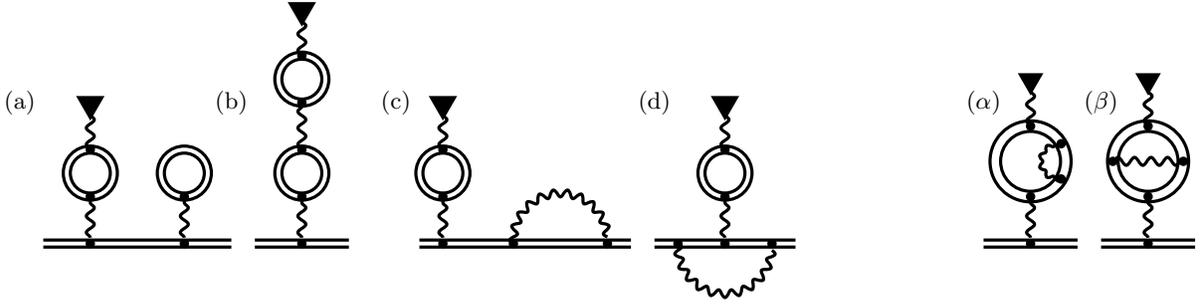
\begin{figure*}[bt]
  \begin{tikzpicture}[very thick,scale=.625]
    \draw (-8-2,-.075) -- (-8+2,-.075);
    \draw (-8-2,.075) -- (-8+2,.075);

    \fill (-8-1,0) circle (.1);
    \fill (-8+1,0) circle (.1);

    \draw[decorate,decoration={snake,amplitude=1.5,segment length=6.25}] (-8-1,.125)  -- (-8-1,1);
    \draw[decorate,decoration={snake,amplitude=1.5,segment length=6.25}] (-8+1,.125)  -- (-8+1,1);

    \fill (-8-1,1) circle (.1);
    \fill (-8+1,1) circle (.1);

    \draw (-8-1,1+.5) circle (1.15/2);
    \draw (-8-1,1+.5) circle (.85/2);
    \draw (-8+1,1+.5) circle (1.15/2);
    \draw (-8+1,1+.5) circle (.85/2);

    \fill (-8-1,2) circle (.1);

    \draw[decorate,decoration={snake,amplitude=1.5,segment length=6.25}] (-8-1,2)  -- (-8-1,2.875);

    \draw[inner sep=2.5] (-8-1,3) node [magnetic]{};

    \draw (-8-2.5,3) node {(a)};
    \draw (-4.5-1,-.075) -- (-4.5+1,-.075);
    \draw (-4.5-1,.075) -- (-4.5+1,.075);

    \fill (-4.5+0,0) circle (.1);
    \fill (-4.5+0,2) circle (.1);

    \draw[decorate,decoration={snake,amplitude=1.5,segment length=6.25}] (-4.5+0,.125)  -- (-4.5+0,1);
    \draw[decorate,decoration={snake,amplitude=1.5,segment length=6.25}] (-4.5+0,4)  -- (-4.5+0,4.875);

    \fill (-4.5+0,1) circle (.1);
    \fill (-4.5+0,3) circle (.1);

    \draw (-4.5+0,1+.5) circle (1.15/2);
    \draw (-4.5+0,1+.5) circle (.85/2);
    \draw (-4.5+0,3+.5) circle (1.15/2);
    \draw (-4.5+0,3+.5) circle (.85/2);

    \fill (-4.5+0,4) circle (.1);

    \draw[decorate,decoration={snake,amplitude=1.5,segment length=6.25}] (-4.5+0,2)  -- (-4.5+0,3);

    \draw[inner sep=2.5] (-4.5+0,5) node [magnetic]{}; 

    \draw (-4.5-1.5,3) node {(b)};
    \draw (-2,-.075) -- (2.5,-.075);
    \draw (-2,.075) -- (2.5,.075);

    \fill (-1.5,0) circle (.1);

    \draw[decorate,decoration={snake,amplitude=1.5,segment length=6.25}] (-1.5,.125)  -- (-1.5,1);

    \fill (-1.5,1) circle (.1);

    \draw (-1.5,1+.5) circle (1.15/2);
    \draw (-1.5,1+.5) circle (.85/2);

    \fill (-1.5,2) circle (.1);

    \draw[decorate,decoration={snake,amplitude=1.5,segment length=6.25}] (-1.5,2)  -- (-1.5,2.875);

    \draw[inner sep=2.5] (-1.5,3) node [magnetic]{};

    \draw[decorate,decoration={complete sines,amplitude=2.75,segment length=5.}] (0,0) arc (180:0:2);

    \fill (0,0) circle (.1);
    \fill (2,0) circle (.1); 

    \draw (-2.5,3) node {(c)};
    \draw (3.5-.5,-.075) -- (3.5+2.5,-.075);
    \draw (3.5-.5,.075) -- (3.5+2.5,.075);

    \fill (3.5+1,0) circle (.1);

    \draw[decorate,decoration={snake,amplitude=1.5,segment length=6.25}] (3.5+1,.125)  -- (3.5+1,1);

    \fill (3.5+1,1) circle (.1);

    \draw (3.5+1,1+.5) circle (1.15/2);
    \draw (3.5+1,1+.5) circle (.85/2);

    \fill (3.5+1,2) circle (.1);

    \draw[decorate,decoration={snake,amplitude=1.5,segment length=6.25}] (3.5+1,2)  -- (3.5+1,2.875);

    \draw[inner sep=2.5] (3.5+1,3) node [magnetic]{};

    \draw[decorate,decoration={complete sines,amplitude=2.75,segment length=5.}] (3.5+0,0) arc (180:360:2);

    \fill (3.5+0,0) circle (.1);
    \fill (3.5+2,0) circle (.1); 

    \draw (3.5-.5,3) node {(d)};
    \draw (11+-1,-.075) -- (11+1,-.075);
    \draw (11+-1,.075) -- (11+1,.075);

    \fill (11+0,0) circle (.1);
    \fill (11+0,2.5) circle (.1);

    \draw[decorate,decoration={snake,amplitude=1.5,segment length=6.25}] (11+0,.125)  -- (11+0,1);

    \fill (11+0,1) circle (.1);
    \fill (11+.75*.866,1.5-.75*.5+.25) circle (.1);
    \fill (11+.75*.866,1.5+.75*.5+.25) circle (.1);

    \draw (11+0,1+.75) circle (3/2*1.15/2);
    \draw (11+0,1+.75) circle (3/2*.85/2);

    \draw[decorate,decoration={complete sines,amplitude=1.5,segment length=3.75}] (11+.75*.866,1.5-.75*.5+.25) arc (270:90:.75);

    \draw[decorate,decoration={snake,amplitude=1.5,segment length=6.25}] (11+0,2.5)  -- (11+0,2.875+.5);
    
    \draw[inner sep=2.25] (11+0,3.5) node [magnetic]{};

  \draw (11+-1,3) node {($\alpha$)};
    \draw (13.5+-1,-.075) -- (13.5+1,-.075);
    \draw (13.5+-1,.075) -- (13.5+1,.075);

    \fill (13.5+0,0) circle (.1);
    \fill (13.5+0,2.5) circle (.1);

    \draw[decorate,decoration={snake,amplitude=1.5,segment length=6.25}] (13.5+0,.125)  -- (13.5+0,1);

    \fill (13.5+0,1) circle (.1);
    \fill (13.5+.75,1.75) circle (.1);
    \fill (13.5+-.75,1.75) circle (.1);

    \draw (13.5+0,1+.75) circle (3/2*1.15/2);
    \draw (13.5+0,1+.75) circle (3/2*.85/2);

    \draw[decorate,decoration={snake,amplitude=1.5,segment length=6.25}] (13.5+-.75,1.75) -- (13.75+.5,1.75);

    \draw[decorate,decoration={snake,amplitude=1.5,segment length=6.25}] (13.5+0,2.5)  -- (13.5+0,2.875+.5);
    
  \draw (13.5+-1,3) node {($\beta$)};
    \draw[inner sep=2.25] (13.5+0,3.5) node [magnetic]{};
 \end{tikzpicture}
  \vspace{-15pt}
    \caption{The diagrams corresponding to the electric-loop--magnetic-loop (a), magnetic-loop-after-loop (b) contributions, and to the wave function-type (c) and vertex-type (d) self-energy--magnetic-loop contributions to the $g$ factor of a bound electron. The double line represents the bound electron, internal wavy lines are intermediate photons, while the wavy line terminated by a triangle denotes a photon from the external magnetic field. The SE-in-ML diagrams ($\alpha$) and ($\beta$) vanish together with the MLAL diagram (b), in the free-loop approximation, due to the properties of the K\"all\'en-Sabry potential.  Diagrams (a), (c) and ($\alpha$) each have an equivalent diagram, therefore, their contributions should be counted twice. \label{fig:FDiagram}}
\end{figure*}

\textit{Virtual light-by-light scattering.---} In a subset of the VP-VP and VP-SE diagrams examined in Ref.~\cite{VladZVP}, the photon from the external magnetic field is attached to a VP loop. This is called the \emph{magnetic loop} (ML), and vanishes in the free VP loop approximation~\cite{DelbrueckGArticle,DelbrueckGLetter}.  All diagrams considered in this work contain a ML \emph{and} another loop, corresponding to either another fermionic pair (electric-loop--magnetic-loop (EL-ML) diagram, see Fig.~\ref{fig:FDiagram}~(a) and magnetic-loop-after-loop (MLAL) diagram, see Fig.~\ref{fig:FDiagram}~(b)) or to a virtual photon (self-energy--magnetic-loop (SE-ML) diagrams, see Figs.~\ref{fig:FDiagram}~(c) and~(d)). To the lowest contributing order, the fermion propagator in the ML interacts twice with the Coulomb field of the nucleus~\cite{BeierOneLoop}, so that Delbr\"uck scattering is a subprocess of the overall diagram. Approximating the contribution of the ML by this light-by-light scattering process has been found to be satisfactory even for highly charged ions~\cite{LeeDelbrueck} in the one-loop case. In what follows, we will make use of the Delbr\"uck-scattered vector potential, which is given~\cite{DelbrueckGArticle} in momentum space by
\begin{equation} \label{eq:DelbMagField}
  A_{\mathrm{ML}\,i}\left(\mathbf{q}\right)=\frac{4\pi}{\mathbf{q}^2}\int\frac{\mathrm{d}\mathbf{k}}{\left(2\pi\right)^3}M_{ji}\left(\mathbf{k},\mathbf{q}\right)A_j\left(\mathbf{k}\right),
\end{equation}
where we sum over repeated indices. Here $\mathbf{A}$ is the external vector potential, given in momentum space by
\begin{equation} \label{eq:ExternalMag}
  \mathbf{A}\left(\mathbf{k}\right)=\frac{\mathrm{i}}{2}\left(2\pi\right)^3\left[\mathbf{B}\times\bm{\nabla}_{\mathbf{k}}\,\delta\left(\mathbf{k}\right)\right]
\end{equation}
and by $\mathbf{A}\left(\mathbf{x}\right)=\left(1/2\right)\left(\mathbf{B}\times\mathbf{x}\right)$ in configuration space, with $\mathbf{B}$ the homogeneous external magnetic field. Also, $M$ is the (tensor) Delbr\"uck scattering amplitude, which is known in closed form in the partial low-energy limit $\left|\mathbf{k}\right|\rightarrow0$ corresponding to a static, homogeneous magnetic field~\cite{LeeDelbrueck,DelbrueckGArticle}, and reads
\begin{equation} \label{eq:DelbrueckTensorRemind}
  M_{ji}\left(\mathbf{k},\mathbf{q}\right)=\alpha\,\lambdabar_e^3\left[\delta_{ji}\left(\mathbf{k}\cdot\mathbf{q}\right)-q_jk_i\right]\left(Z\alpha\right)^2F_{\mathrm{D}}\left(\lambdabar_e\left|\mathbf{q}\right|\right),
\end{equation}
with the reduced Compton wavelength $\lambdabar_e=\hbar/m_e c$. The explicit expression for the Delbr\"uck scattering function $F_{\mathrm{D}}$ is given in Ref.~\cite{LeeDelbrueck}. It is also helpful to write the expression of the configuration-space Delbr\"uck-scattered vector potential:
\begin{equation} \label{eq:DelbMagFieldConfig}
  \mathbf{A}_{\mathrm{ML}}\left(\mathbf{x}\right)=\frac{1}{2}\left(\mathbf{B}\times\mathbf{x}\right)\Pi_{\mathrm{ML}}\left(\frac{\left|\mathbf{x}\right|}{\lambdabar_e}\right),
\end{equation}
with the polarization function given by a Bessel transform of the scattering amplitude:
\begin{equation} \label{eq:PiPolar}
  \Pi_{\mathrm{ML}}\left(u\right)=4\frac{\alpha}{\pi}\left(Z\alpha\right)^2\frac{1}{u^2}\int_0^{+\infty}\mathrm{d}z\,zu\,j_1\left(zu\right)F_{\mathrm{D}}\left(z\right).
\end{equation}
The Delbr\"uck-scattered vector potential (\ref{eq:DelbMagFieldConfig}) thus has the same angular structure as the external vector potential.\\
\textit{Calculations.---} The contribution from the electric-loop--magnetic-loop diagram of Fig.~\ref{fig:FDiagram}(a) to the $g$ factor of the bound electron can be deduced from the simpler ML diagram studied in Refs.~\cite{DelbrueckGLetter,DelbrueckGArticle,LeeDelbrueck}. For the ML diagram, it can be shown, for instance by using the two-time Green's function formalism~\cite{Shabaev}, that the correction to the energy $E_a$ of level $a$ is given by
\begin{equation} \label{eq:MaLoop}
  \Delta E_a^{\mathrm{ML}}=\int\frac{\mathrm{d}\mathbf{k}}{\left(2\pi\right)^3}\mathbf{A}_{\mathrm{ML}}\left(\mathbf{k}\right)\cdot\mathbf{j}_a^*\left(\mathbf{k}\right),
\end{equation}
where $\mathbf{j}_a$ is the Dirac current in the reference state $a$ (taken to be the ground state $1s$ throughout this work) of the bound electron. The inclusion of the EL is then performed by amending the electronic current in Eq.~(\ref{eq:MaLoop}). The EL does not modify the angular structure of the wave functions. Accordingly, we write the EL-ML correction to the $g$ factor as~\cite{DelbrueckGArticle,LeeDelbrueck}
\begin{multline} \label{eq:ElMaLoopRadial}
  \Delta g_a^{\mathrm{EL-ML}}=-\frac{8}{3}\frac{1}{\lambdabar_e}\int_0^{+\infty}\mathrm{d}r\,\,r^3\,\Pi_{\mathrm{ML}}\left(\frac{r}{\lambdabar_e}\right)\\
  \times\left[g_a\left(r\right)\delta_{\mathrm{VP}}f_a\left(r\right)+\delta_{\mathrm{VP}}g_a\left(r\right)f_a\left(r\right)\right],
\end{multline}
where $g_a$ and $f_a$ are the radial parts of the Dirac-Coulomb spinor $\psi_a$~\cite{ForVertexF}. The EL potential is approximated~\cite{BeierOneLoop} by the Uehling potential and the Uehling-corrected radial wave functions are $\delta_{\mathrm{VP}}g_a$ and $\delta_{\mathrm{VP}}f_a$. For this diagram, the radial wave functions are computed numerically for finite-size nuclei. To achieve this, the electron is confined in a radial cavity and the dual kinetic balance approach~\cite{Shabaev2004} is used to generate numerical spectra and radial wave functions. In our calculations, the nucleus is considered to be a homogeneously charged sphere and the resulting Uehling potential is computed using analytical expressions given in Ref.~\cite{Klarsfeld1977}.

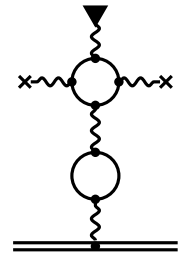
\begin{figure}[b!]
  \begin{tikzpicture}[very thick,scale=.625]
    \draw (-1.75,-.075) -- (1.75,-.075);
    \draw (-1.75,.075) -- (1.75,.075);

    \fill (0,0) circle (.1);
    \fill (0,2) circle (.1);

    \draw[decorate,decoration={snake,amplitude=1.5,segment length=6.25}] (0,.125)  -- (0,1);
    \draw[decorate,decoration={snake,amplitude=1.5,segment length=6.25}] (0,4)  -- (0,4.875);

    \fill (0,1) circle (.1);
    \fill (0,3) circle (.1);

    \draw (0,1+.5) circle (1/2);
    \draw (0,3+.5) circle (1/2);

    \fill (0,4) circle (.1);
    \fill (-.5,3.5) circle (.1);
    \fill (.5,3.5) circle (.1);

    \draw[decorate,decoration={snake,amplitude=1.5,segment length=6.25}] (-.5,3.5)  -- (-1.375,3.5);
    \draw[decorate,decoration={snake,amplitude=1.5,segment length=6.25}] (.5,3.5)  -- (1.375,3.5);

    \draw (-1.5-.125,3.5-.125) -- (-1.5+.125,3.5+.125);
    \draw (-1.5-.125,3.5+.125) -- (-1.5+.125,3.5-.125);
    \draw (1.5-.125,3.5-.125) -- (1.5+.125,3.5+.125);
    \draw (1.5-.125,3.5+.125) -- (1.5+.125,3.5-.125);

    \draw[decorate,decoration={snake,amplitude=1.5,segment length=6.25}] (0,2)  -- (0,3);

    \draw[inner sep=2.25] (0,5) node [magnetic]{};

  \end{tikzpicture}
   \caption{The ``scarecrow'' diagram represents the lowest nonvanishing term that contributes to the magnetic loop-after-loop diagram given in Fig.~\ref{fig:FDiagram}(b). \label{fig:Scarecrow}}
\end{figure}
The contribution from the magnetic-loop-after-loop diagram of Fig.~\ref{fig:FDiagram}(b) is computed by another modification to the ML diagram. At the free-loop level, the MLAL diagram vanishes, along with the SE-in-ML diagrams (see Fig.~\ref{fig:FDiagram}($\alpha$) and ($\beta$))~\cite{VladZVP}. This is due to the low-momentum properties of the (K\"all\'en-Sabry) fourth-order VP tensor~\cite{KaellenSabry}, in the same way that the vanishing of the simpler one-loop ML diagram at the free-loop level~\cite{BeierOneLoop,DelbrueckGArticle} is due to the low-momentum properties of the (Uehling) second-order VP tensor. Hence, the lowest nonvanishing contribution from the SE-in-ML diagrams features a six-photon light-by-light scattering process, and is out of the scope of the present work. The first nonvanishing contribution from the MLAL diagram comes from adding two interactions with the Coulomb field of the nucleus, on the outermost free loop, while keeping the innermost loop free (we call this the scarecrow diagram, see Fig.~\ref{fig:Scarecrow}). The diagram wherein the Coulomb photons interact with the innermost loop and the outermost loop is free, has a vanishing contribution. Hence at the lowest nonvanishing order in the VP loops, the contribution of the MLAL diagram is given by
\begin{equation} \label{eq:MLALDiagram}
  \Delta g_a^{\mathrm{MLAL}}=-\frac{8}{3}\frac{1}{\lambdabar_e}\int_0^{+\infty}\mathrm{d}r\,\,r^3\,\Pi_{\mathrm{MLAL}}\left(\frac{r}{\lambdabar_e}\right)f_a\left(r\right)g_a\left(r\right),
\end{equation}
where the photon interacting with the bound electron has gone through two VP loops and is described by the MLAL polarization function
\begin{multline} \label{eq:PiDoublePolar}
    \Pi_{\mathrm{MLAL}}\left(u\right)=4\left(\frac{\alpha}{\pi}\right)^2\left(Z\alpha\right)^2\frac{1}{u^2}\\
    \times\int_0^{+\infty}\mathrm{d}z\,zu\,j_1\left(zu\right)F_{\mathrm{D}}\left(z\right)I_{\mathrm{VP}}\left(z\right).
\end{multline}
Here the one-loop photon VP function is given by~\cite{VladZVP}
\begin{equation} \label{eq:VPLoop}
  I_{\mathrm{VP}}\left(z\right)=z^2\int_0^1\mathrm{d}\tau\frac{\tau^2\left(1-\frac{\tau^2}{3}\right)}{4+\left(1-\tau^2\right)z^2}.
\end{equation}
The MLAL contribution (\ref{eq:MLALDiagram}) to the $g$ factor is then computed for pointlike nuclei.

The contribution from the wave function-type self-energy--magnetic-loop diagram of Fig.~\ref{fig:FDiagram}(c) can be split into a reducible contribution, wherein the intermediate state in the bound electron propagator between the SE loop and the ML is the ground state $a=1s$, and an irreducible contribution, wherein a sum over all intermediate states, excluding the ground state, is performed in that propagator. The irreducible contribution $\Delta g_{a\left(\mathrm{irr}\right)}^{\mathrm{SE-ML}}$ is computed in the same way as the EL-ML diagram: as is the case of the EL, the SE loop does not modify the angular dependence of the wave function of the bound electron, so that, if in Eq.~(\ref{eq:ElMaLoopRadial}), the VP-corrected wave functions are replaced with the SE-corrected wave functions $\delta_{\mathrm{SE}}f_a$ and $\delta_{\mathrm{SE}}g_a$, we straightforwardly obtain the irreducible correction to the $g$ factor due to the SE-ML diagram. The irreducible contribution was computed with SE-corrected wave functions obtained with the method presented in Ref.~\cite{SelfEnergyWFNat}. The reducible contribution, on the other hand, is given by
\begin{equation} \label{eq:Proport}
  \Delta g_{a\left(\mathrm{red}\right)}^{\mathrm{SE-ML}}=\frac{\Delta g_a^{\mathrm{ML}}}{g_{a\left(\mathrm{D}\right)}}\Delta g_{a\left(\mathrm{red}\right)}^{\mathrm{SE}}
\end{equation}
where $g_{a\left(\mathrm{D}\right)}$ is simply the Breit-Dirac value for the bound-electron $g$ factor~\cite{BreitG}. Both the reducible contribution to the one-loop SE correction $\Delta g_{a\left(\mathrm{red}\right)}^{\mathrm{SE}}$~\cite{YeroGFactor} and the ML correction $\Delta g_a^{\mathrm{ML}}$~\cite{DelbrueckGArticle} to the $g$ factor have been investigated previously, therefore, we can directly compute the reducible contribution (\ref{eq:Proport}) to the two-loop correction for pointlike nuclei. There is a potential further reducible contribution to the SE-ML diagram, coming from the energy derivative of the electron propagators in the ML, which vanishes because these free propagators do not depend on the energy of the bound electron.

Let us finally turn to the contribution from the vertex SE-ML diagram of Fig.~\ref{fig:FDiagram}(d). As was the case for the simpler vertex SE one-loop correction~\cite{YeroGFactor}, we need, for renormalization purposes, to split this diagram into a zero-potential part (whereby the propagator of the electron under the SE loop is taken to be that of the free electron) and a many-potential part (whereby the electron interacts any nonzero number of times with the Coulomb field of the nucleus under the SE loop). The zero-potential term can be treated analytically to a large extent. After renormalization, its general expression is given by
\begin{multline} \label{eq:VertexImplicit}
  \Delta g_{a\left(\mathrm{ver}\right)}^{\mathrm{SE-ML}\left(0\right)}=\frac{2}{\mu_a\,\lambdabar_e\,B}\int\frac{\mathrm{d}\mathbf{p}}{\left(2\pi\right)^3}\int\frac{\mathrm{d}\mathbf{p}'}{\left(2\pi\right)^3}\bar{\psi}_a\left(\mathbf{p}\right)\\\times\bm{\Gamma}_R\left(p,p'\right)\cdot\mathbf{A}_{\mathrm{ML}}\left(\mathbf{p}-\mathbf{p}'\right)\psi_a\left(\mathbf{p}'\right),
\end{multline}
with $\mu_a$ the magnetic projection quantum number, and the Delbr\"uck-scattered vector potential $\mathbf{A}_{\mathrm{ML}}$ given by Eq.~(\ref{eq:DelbMagField}). The four-vectors $p$ and $p'$ share the same time component fixed by the energy $\epsilon_a$ of the reference state: $p=\left(\epsilon_a/c,\mathbf{p}\right)$, $p'=\left(\epsilon_a/c,\mathbf{p}'\right)$, while $\mathbf{\Gamma}_R$ is the UV-finite part of the free vertex function, studied in detail in Ref.~\cite{ForVertexF}. Performing three of the four angular integrals in Eq.~(\ref{eq:VertexImplicit}) analytically, the zero-potential term in the vertex diagram can be cast as a quadruple integral (a double radial integral, one remaining angular integral, and an integral over a Feynman parameter present from the outset in $\Gamma_R$), to be performed numerically for pointlike nuclei. The many-potential term, on the other hand, is computed in configuration space, and is treated in a very similar way to the corresponding term in the one-loop SE correction~\cite{YeroGFactor,BastianPhD}: the bound and free electronic Green's function are expanded in partial waves according to the absolute value $\left|\kappa\right|$ of the Dirac angular momentum. As can be seen from Eq.~(\ref{eq:DelbMagFieldConfig}), the angular structure of the Delbr\"uck-scattered vector potential (\ref{eq:DelbMagField}) is identical, in configuration space, to that of the external vector potential (\ref{eq:ExternalMag}), meaning that the only modification to the calculation concerns the radial integrals. As was the case for the one-loop SE correction, the many-potential term in the vertex diagram features an infrared divergence, that is rigorously cancelled by a divergence in the many-potential, reducible contribution~\cite{YeroGFactor} (in contrast to the approach of Ref.~\cite{YeroGFactor}, we do not separate the (finite) one-potential term here). We computed the sum of the many-potential terms of the reducible and vertex contributions to the SE-ML correction, which is readily finite because of that cancellation. The partial wave summation is truncated at $\left|\kappa\right|=20$, with the remaining terms estimated through least-squares inverse polynomial fitting. 

\begin{table*}[t!]
\begin{center}
\begin{minipage}{1\linewidth}
\begin{center}
\caption{Numerical values of the two-loop magnetic-loop correction $\Delta g_a^{\left(\mathrm{2L-ML}\right)}$ to the $g$ factor of hydrogenlike ions ($a=1s$) for several nuclear charges $Z$. We give separately the contribution $\Delta g_a^{\mathrm{EL-ML}}$ from the electric-loop--magnetic-loop diagram, the contribution $\Delta g_a^{\mathrm{MLAL}}$ from the magnetic-loop-after-loop diagram, and the irreducible $\Delta g_{a\left(\mathrm{irr}\right)}^{\mathrm{SE-ML}}$, zero-potential vertex+reducible $\Delta g_{a\left(\mathrm{ver+red}\right)}^{\mathrm{SE-ML\left(0\right)}}$ and many-potential vertex+reducible $\Delta g_{a\left(\mathrm{ver+red}\right)}^{\mathrm{SE-ML}\left(1+\right)}$ contributions from the SE-ML diagrams. The total correction is given in the last column. Results are given in units of $10^{-6}$, and powers of $10$ are given between square brackets. \label{tab:Res}} 
\begin{tabular}{c@{\hskip 0.05in}|@{\hskip 0.05in}r@{\hskip 0.275in}r@{\hskip 0.275in}r@{\hskip 0.275in}r@{\hskip 0.275in}r@{\hskip 0.05in}|@{\hskip 0.05in}r}
\hline
\hline
\rule[-3mm]{0mm}{8mm}
$Z$ & \multicolumn{1}{c}{$\Delta g_a^{\mathrm{EL-ML}}$} & \multicolumn{1}{c}{$\Delta g_a^{\mathrm{MLAL}}$} & \multicolumn{1}{c}{$\Delta g_{a\left(\mathrm{irr}\right)}^{\mathrm{SE-ML}}$} & \multicolumn{1}{c}{$\Delta g_{a\left(\mathrm{ver+red}\right)}^{\mathrm{SE-ML}\left(0\right)}$} & \multicolumn{1}{c}{$\Delta g_{a\left(\mathrm{ver+red}\right)}^{\mathrm{SE-ML}\left(1+\right)}$} & \multicolumn{1}{c}{$\Delta g_a^{\left(\mathrm{2L-ML}\right)}$}\\
\hline
\rule[-3mm]{0mm}{8mm}
$1$\ & $3.9(1)\hfill\left[-14\right]$ & $3.7\hfill\left[-14\right]$ & $-8.6(8)\hfill\left[-14\right]$ & $\hphantom{-}3.0928(1)\hfill\left[-11\right]$ & $-2.600(250)\hfill\left[-11\right]$ & $4.92(2.50)\hfill\left[-12\right]$\\
\rule[-3mm]{0mm}{6mm}
$2$\ & $2.58(1)\hfill\left[-12\right]$ & $2.34\hfill\left[-12\right]$ & $-8.9(5)\hfill\left[-12\right]$ & $\hphantom{-}9.4430(8)\hfill\left[-10\right]$ & $-7.926(6)\hfill\left[-10\right]$ & $1.477(8)\hfill\left[-10\right]$\\
\rule[-3mm]{0mm}{6mm}
$14$\ & $3.60(1)\hfill\left[-7\right]$ & $2.42\hfill\left[-7\right]$ & $-1.920(8)\hfill\left[-6\right]$ & $\hphantom{-}9.2140(1)\hfill\left[-6\right]$ & $-8.856(6)\hfill\left[-6\right]$ & $-9.60(10)\hfill\left[-7\right]$\\
\rule[-3mm]{0mm}{6mm}
$20$\ & $3.189(3)\hfill\left[-6\right]$ & $1.941\hfill\left[-6\right]$ & $-1.662(3)\hfill\left[-5\right]$ & $\hphantom{-}4.2219(1)\hfill\left[-5\right]$ & $-4.457(3)\hfill\left[-5\right]$ & $-1.384(4)\hfill\left[-5\right]$\\
\rule[-3mm]{0mm}{6mm}
$54$\ & $1.4344(26)\hfill\left[-3\right]$ & $6.019(1)\hfill\left[-4\right]$ & $-5.3135(50)\hfill\left[-3\right]$ & $\hphantom{-}1.0331(1)\hfill\left[-3\right]$ & $-3.3842(24)\hfill\left[-3\right]$ & $-5.628(6)\hfill\left[-3\right]$\\
\rule[-3mm]{0mm}{6mm}
$82$\ & $2.0982(8)\hfill\left[-2\right]$ & $6.845(3)\hfill\left[-3\right]$ & $-5.5379(20)\hfill\left[-2\right]$ & $-4.0125(94)\hfill\left[-3\right]$ & $-2.0833(22)\hfill\left[-2\right]$ & $-8.851(3)\hfill\left[-2\right]$\\
\rule[-3mm]{0mm}{6mm}
$92$\ & $4.5676(38)\hfill\left[-2\right]$ & $1.3648(11)\hfill\left[-2\right]$ & $-1.0570(5)\hfill\left[-1\right]$ & $-1.1957(43)\hfill\left[-2\right]$ & $-3.4565(31)\hfill\left[-2\right]$ & $-9.290(8)\hfill\left[-2\right]$\\
\hline
\hline
\end{tabular}
\end{center}
\end{minipage}
\end{center}
\end{table*}
\textit{Results.---} We present numerical results for seven specific hydrogen-like ions, with the nuclear charges $Z=1,\,2,\,14,\,20,\,54,\,82,\,92$. The contributions from all diagrams are summarized in Table~\ref{tab:Res}.
\begin{figure}[b!]
  \begin{center}
    \includegraphics[width=.475\textwidth]{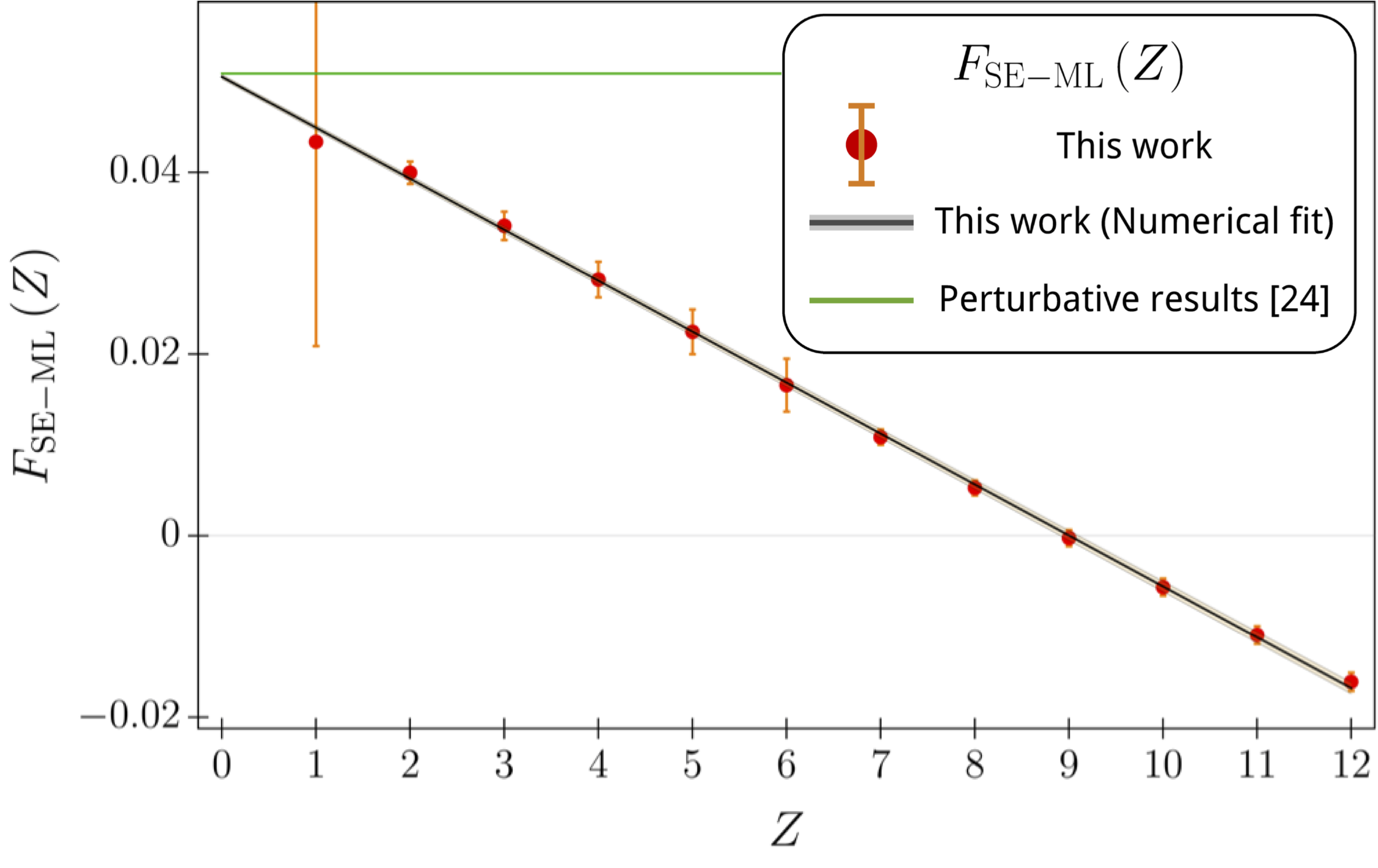}
  \end{center}
  \caption{Numerical values (red dots) of the SE-ML contribution to the $g$ factor of hydrogenlike ions for low nuclear charges, as parametrized through Eq.~(\ref{eq:Param}). The error bars (orange) have been artificially magnified by a factor of $4$ in order to be visible, except for $Z=1$ where the actual error bar is shown. Our results are fitted (black line) according to Eq.~(\ref{eq:Fit}), with the shaded gray region around the line corresponding to the uncertainty on the fit parameters. The agreement with the perturbative result (green line) of Ref.~\cite{CzarneckiLetter} is manifest. However, it is seen that the SE-ML contribution markedly departs from the prediction of Ref.~\cite{CzarneckiLetter} even at very low $Z$. \label{fig:LowZ}}
\end{figure}  

The EL-ML and irreducible SE-ML corrections were computed for finite-size nuclei, with the nuclear radii \footnote{Since the nuclear radii also depend on the isotope considered, we clarify here that we use the most abundant isotopes $_1^1\text{H}$, $_2^4\text{He}^{+}$, $_{14}^{28}\text{Si}^{13+}$, $_{20}^{40}\text{Ca}^{19+}$, $_{54}^{132}\text{Xe}^{53+}$, $_{82}^{208}\text{Pb}^{81+}$, $_{92}^{238}\text{U}^{91+}$. In the case of xenon, the isotope used here has the radius which is known with the best accuracy~\cite{NuclRadiiUpdate}.} taken from Ref.~\cite{NuclRadiiUpdate}. The MLAL correction, as well as the reducible and vertex SE-ML corrections, were computed for pointlike nuclei. The uncertainty of the EL-ML correction is dominated by that on the nuclear radii. The uncertainty of the MLAL correction is given by the estimated finite-nuclear-size correction. The uncertainty of the SE-ML correction is dominated by numerical convergence and, for high $Z$ ($Z=54$, $Z=82$ and $Z=92$, see Table~\ref{tab:Res}), by nuclear size corrections.

As anticipated for instance in Ref.~\cite{VladZVP}, and as was confirmed in Ref.~\cite{CzarneckiLetter} in the perturbative approach, all corrections obtained here are reliably smaller than the SE-VP~\cite{VladZVP} and VP-VP~\cite{SikoraTwoLoop} corrections from which the ML is absent, especially at low and intermediate $Z$. Nevertheless, for high $Z$, the calculated corrections are large enough to be above the experimental uncertainties of state-of-the-art measurements of the $g$ factor of bound electrons. Indeed the uncertainty reported in the most precise measurements of the $g$ factor~\cite{Sturm13,Sturm14} is of the order of $10^{-11}$. Although these measurements were only performed on lighter ions such as carbon ($Z=6$) and silicon ($Z=14$), it is expected that, in the framework of the ALPHATRAP project~\cite{Sturm19}, $g$-factor measurements on heavier ions will be performed at comparable levels of accuracy, meaning that the contributions computed here should be taken into account in an accurate interpretation of these experiments.

At the opposite end of the nuclear charge landscape, we note that our results for the SE-ML contribution approach the prediction of the perturbative approach on Ref.~\cite{CzarneckiLetter} for $Z\rightarrow0$, but that these perturbative results have limited relevance due to large higher-order contributions, which are not captured by the approach of Ref.~\cite{CzarneckiLetter}. This is shown in Fig.~\ref{fig:LowZ}, where our results for low $Z$ for the SE-ML diagrams (Figs.~\ref{fig:FDiagram} (c) and (d)) are parametrized according to
\begin{equation} \label{eq:Param}
  \Delta g^{\mathrm{SE-ML}}=\left(\frac{\alpha}{\pi}\right)^2\left(Z\alpha\right)^5F_{\mathrm{SE-ML}}\left(Z\right).
\end{equation}
In Ref.~\cite{CzarneckiLetter} the leading term in the function $F_{\mathrm{SE-ML}}\left(Z\right)$ is given by $F_{\mathrm{SE-ML}}\left(Z\right)=\left(7/432\right)\pi\simeq0.0509$. However, we find that a linear term is necessary to capture the behavior of the SE-ML correction to the $g$ factor, even at very low $Z$:
\begin{equation} \label{eq:Fit}
  F_{\mathrm{SE-ML}}\left(Z\right)=a_5+a_6\left(Z\alpha\right),
\end{equation}
with the values of the coefficients $a_5=0.0505(3)$ and $a_6=-0.769(4)$ obtained through a least-squares numerical fit. This confirms the result of Ref.~\cite{CzarneckiLetter}. However, the large value of the $a_6$ coefficient causes the corresponding term, which brings a correction to the $g$ factor proportional to $\left(Z\alpha\right)^6$, to be important already for $Z<5$. At $Z=8$--$10$, a sign flip occurs.

\textit{Discussion.---} The corrections computed here are several orders of magnitude smaller than the leading nuclear corrections to the $g$ factor, namely, the leading-order nuclear recoil (finite mass) correction~\cite{RecoilAllOrders} and the leading-order finite nuclear size correction~\cite{ShabaevSize}. They are broadly comparable to subdominant nuclear corrections: a higher-order contribution to the finite nuclear size correction~\cite{FNSGHigher}, the nuclear polarizability~\cite{NuPolAna,NuPolNum} and deformation corrections~\cite{NuDefJacek}, and the higher-order mass correction~\cite{HOMass}. Importantly, the two-loop corrections computed here are of the same order of magnitude as the uncertainty in the finite nuclear size correction, as can be checked from Refs.~\cite{NuclRadiiUpdate} and~\cite{ShabaevSize}. As a result, they are arguably mostly relevant within the framework of the `specific' weighted difference between the $g$-factor of H-like and Li-like ions, which allows for the approximate cancellation of nuclear size corrections~\cite{WDiffOld,GFactorAlpha,WDiffZ}.

\textit{Conclusion.---} Two-loop QED corrections to the bound electron $g$ factor involving the magnetic loop were calculated for the first time in a nonperturbative approach. The calculated corrections deviate significantly from the perturbative results~\cite{CzarneckiLetter} and are substantially larger than projected experimental uncertainties for heavy hydrogenlike ions, of relevance for tests of QED, searches for New Physics, and the determination of fundamental constants.

The authors thank Krzysztof Pachucki for helpful discussions. This work is funded by the Deutsche Forschungsgemeinschaft (DFG, German Research Foundation) -- Project-ID 273811115 -- SFB 1225 (ISOQUANT).

\bibliography{Biblio}
\end{document}